\documentclass[twocolumn,showpacs,preprintnumbers,amsmath,amssymb]{revtex4-1}
\usepackage{graphicx}
\usepackage{rotating}

\usepackage{times}
\usepackage{amsmath}
\usepackage{amssymb}
\usepackage {amsthm}
\usepackage {longtable}
\usepackage[usenames,dvipsnames]{color}
\usepackage[utf8]{inputenc}

\begin{document}

\renewcommand{\theequation}{\thesection.\arabic{equation}}

\newcommand{\re}{\mathop{\mathrm{Re}}}

\newcommand{\be}{\begin{equation}}
\newcommand{\ee}{\end{equation}}
\newcommand{\bea}{\begin{eqnarray}}
\newcommand{\eea}{\end{eqnarray}}

%\maketitle

\title{Varying constants quantum cosmology}

\author{Katarzyna Leszczy\'nska}
\email{k.leszczynska@wmf.univ.szczecin.pl}
\affiliation{\it Institute of Physics, University of Szczecin, Wielkopolska 15, 70-451 Szczecin, Poland}

\author{Adam Balcerzak}
\email{abalcerz@wmf.univ.szczecin.pl}
\affiliation{\it Institute of Physics, University of Szczecin, Wielkopolska 15, 70-451 Szczecin, Poland}
\affiliation{\it Copernicus Center for Interdisciplinary Studies, S{\l }awkowska 17, 31-016 Krak\'ow, Poland}

\author{Mariusz P. D\c{a}browski}
\email{mpdabfz@wmf.univ.szczecin.pl}
\affiliation{\it Institute of Physics, University of Szczecin, Wielkopolska 15, 70-451 Szczecin, Poland}
\affiliation{\it Copernicus Center for Interdisciplinary Studies,
S{\l }awkowska 17, 31-016 Krak\'ow, Poland}

\date{\today}

\input epsf

\begin{abstract}
We discuss minisuperspace models within the framework of varying physical constants theories including $\Lambda$-term. In particular, we consider the varying speed of light (VSL) theory and varying gravitational constant theory (VG) using the specific ans\"atze for the variability of constants: $c(a) = c_0 a^n$ and $G(a)=G_0 a^q$. We find that most of the varying $c$ and $G$ minisuperspace potentials are of the tunneling type which allows to use WKB approximation of quantum mechanics. Using this method we show that the probability of tunneling of the universe ``from nothing'' ($a=0)$ to a Friedmann geometry with the scale factor $a_t$ is large for growing $c$ models and is strongly suppressed for diminishing $c$ models. As for $G$ varying, the probability of tunneling is large for $G$ diminishing, while it is small for $G$ increasing. In general, both varying $c$ and $G$ change the probability of tunneling in comparison to the standard matter content (cosmological term, dust, radiation) universe models.
\end{abstract}


\maketitle

\section{Introduction}
\label{intro}
\setcounter{equation}{0}

The early idea of variation of physical constants \cite{varconst} has been established widely in physics both theoretically and experimentally \cite{uzan}. The gravitational constant $G$, the charge of electron $e$, the velocity of light $c$, the proton to electron mass ratio $\mu = m_p/m_e$, and the fine structure constant $\alpha = e^2/\hbar c$, where $\hbar$ is the Planck constant, may vary in time and space \cite{barrowbook}. The earliest and best-known framework for varying $G$ theories has been Brans-Dicke theory \cite{bd}. Nowadays, the most popular theories which admit physical constants variation are the varying $\alpha$ theories \cite{alpha}, varying gravitational constant $G$ theories \cite{VG}, and the varying speed of light $c$ theories \cite{UzanLR,VSL} (see also a critical discussion on the meaning of the variation of $c$ in Ref. \cite{EllisUzan03}). The latter theory allows the solution of the standard cosmological problems such as the horizon problem, the flatness problem, the $\Lambda-$problem, and has recently been proposed to solve the singularity problem \cite{JCAP13}. All these considerations were devoted to classical varying constants cosmology.

The first attempt of varying constants quantum cosmology was given in Ref. \cite{harko2000} in which both $c$ and $G$ as well as the cosmological constant $\Lambda$ were considered to change. The probability of tunneling from nothing to a de Sitter phase was calculated - though under some specific assumption related to the energy conservation equation. In Ref. \cite{szydloQC} only variability of $c$ was taken into account. The claim was that the probability of tunneling was largest for a constant value of the speed of light. However, the result was obtained under admitting the decrease of the speed of light only which was based on the previous observations \cite{webb99,murphy2007}. Nowadays, both options - the decrease and increase of $c$ are reported \cite{webb2011,king2012}, so that one should not restrict oneself to such a case only. In fact, for speed of light increasing, the probability of tunneling is monotonically increasing either and no maximum appears. In yet another Ref. \cite{yurov2} more minisuperspace models for varying $c$ were discussed, but with some restrictive assumption related to matter content.

Our paper is organized as follows. In Sec. \ref{VC} we formulate the basics of the classical varying $c$ and varying $G$ theories which generalize Einstein's general relativity. In Sec. \ref{Action} we obtain minisuperspace Wheeler-DeWitt equation starting from generalized Einstein-Hilbert action of general relativity. In Sec. \ref{tunnel} we derive the probability of tunneling for varying constants cosmology using special ansatz for the variability of the speed of light $c(a) = c_0 a^n$ and gravitational constant $G(a) = G_0 a^q$, where $a$ is the scale factor and $c_0$, $G_0$, $n$ and $q$ are constants. In Sec. \ref{conclusion} we give our conclusions.

\section{Classical varying constants cosmology}
\setcounter{equation}{0}
\label{VC}

Following Ref. \cite{VG,UzanLR,VSL}, we consider the Friedmann universes within the framework of varying speed of light theories (VSL) and varying gravitational (VG) constant theories. The field equations read as
\bea \label{rho} \varrho(t) &=& \frac{3}{8\pi G(t)}
\left(\frac{\dot{a}^2}{a^2} + \frac{kc^2(t)}{a^2}
\right)~,\\
\label{p} p(t) &=& - \frac{c^2(t)}{8\pi G(t)} \left(2 \frac{\ddot{a}}{a} + \frac{\dot{a}^2}{a^2} + \frac{kc^2(t)}{a^2} \right)~,
\eea
and the energy-momentum conservation law is
\be
\label{conser}
\dot{\varrho}(t) + 3 \frac{\dot{a}}{a} \left(\varrho(t) + \frac{p(t)}{c^2(t)} \right) = - \varrho(t) \frac{\dot{G}(t)}{G(t)}
+ 3 \frac{kc(t)\dot{c}(t)}{4\pi Ga^2}~.
\ee
Here $a \equiv a(t)$ is the scale factor, $p$ is the pressure, $\varrho$ is the mass density, the dot means the derivative with respect to the cosmic time $t$, $G=G(t)$ is time-varying gravitational constant, $c=c(t)$ is time-varying speed of light, and the curvature index $k=0, \pm 1$. Further, it will also be useful to apply the energy density which is $\varepsilon = \varrho c^2$.

If one adds the $\Lambda$-term to the equations (\ref{rho})-(\ref{conser}), and introduces the vacuum mass density
\be
\varrho_{\Lambda}(t) = \frac{\Lambda c^2(t)}{8 \pi G(t)} \hspace{0.5cm} (\Lambda = {\rm const.})
\ee
with
\be
p_{\Lambda}(t) = - \varrho_{\Lambda}(t)c^2,
\ee
then one has to replace $\varrho \to \varrho + \varrho_{\Lambda}$, $p \to p + p_{\Lambda}$ in (\ref{rho})-(\ref{p}) and $\dot{\varrho} \to \dot{\varrho} + \dot{\varrho}_{\Lambda}$ in (\ref{conser}) to obtain
\begin{equation}
\label{conserLam}
\dot{\varrho}+3\frac{\dot{a}}{a}\left(\varrho+\frac{p}{c^2(t)} \right)+ \varrho\frac{\dot{G}(t)}{G(t)}= \frac{\left(3k-\Lambda a^2 \right)}{4\pi G(t)a^2}c(t)\dot{c}(t).
\end{equation}
It is worth noting that the term $(\Lambda c^2 \dot{G})/(8 \pi G^2)$ was canceled on both sides of the equation (\ref{conserLam}). This differs from the derivation presented in the first paper of Ref. \cite{VG} where this term shows up included into the term $\dot{\varrho_{\Lambda}}$ of its Eq. (9). Another point is that the field equations (\ref{rho})-(\ref{conser}) and (\ref{conserLam}) have been obtained from the variational principle which applies in some preferred (minimally coupled) frame in which neither new terms in the Einstein equations nor dynamical equations for $c$ and $G$ treated as scalar fields have been introduced. Though this approach is more disputable for $c-$varying models (cf. Ref. \cite{EllisUzan03}), it has a firm counterpart basis for $G-$varying universes within the framework of Brans-Dicke theory \cite{bd}. However, the appropriate field equations differ from ours (\ref{conser}) and (\ref{conserLam}) in this approach - they do not allow for $G$-varying term $\varrho \dot{G}/G$. Instead, $G$ is replaced by a field $\phi \sim 1/G$ in the field equations which are also appended by the dynamical equation for $\phi$ (see Eqs. (32)-(35) in Ref. \cite{VG}). In our paper we will study quantum cosmology models in the above mentioned minimal coupling approach leaving the matter of a fully dunamical treatment of both $c$ and $G$ fields for a separate work \cite{adam}. A deeper discussion of particular choices of the varying speed of light theories in the context of deriving them from some actions using the standard variational principle was given in Ref. \cite{EllisUzan03}.

It is also worth noticing that the conservation equation (\ref{conserLam}) in Ref. \cite{szydloQC} contains some typos - its Eq. (18) does not contain the term $-\Lambda c \dot{c}/ 4 \pi G$ on the right-hand side and also $\dot{a}/a$ should be replaced by $\dot{c}/c$ in the last term (with the curvature index $K$).

The conservation equation (\ref{conserLam}) is solved by
\bea
\varrho(a) &=& \frac{C}{a^{3(w+1)+q}}  \\
&+& \frac{3c_0^2n}{4\pi G_0}\left(\frac{k}{2n+3w+1}-\frac{\Lambda}{3}\frac{a^2}{2n+3w+3} \right)a^{2(n-1)-q} , \nonumber
\label{varrho}
\eea
where $C=$const.

In this paper we will follow the ansatz for the speed of light given in Ref. \cite{VSL}, i.e.,
\be
\label{c(t)}
c(t) = c_0 a^n(t)~~,
\ee
with the constant speed of light limit $n \to 0$ giving $c(t) \to c_0$. A better ansatz which gives $c_0$ the interpretation of the current value of the speed of light is $c(t) = c_0 (a/a_0)^n$, where $a_0$ is the current value of the scale factor. However, most of the papers nowadays normalize the present value of the scale factor to $a_0 = 1$, and so we have the same result quantitatively. The ansatz for varying gravitational constant will be \cite{VG}
\be
\label{G(t)}
G(t) = G_0 a^q(t)~~.
\ee
Another important point is (see also the full discussion in Ref. \cite{magueijoDer}) that the derivation of Eqs. (\ref{rho})-(\ref{conser}) requires the usage of the coordinate $x^0 = ct$ rather than solely the coordinate $t$. This is because $c=c(t)$ and taking for example partial derivatives of the type $\partial/ \partial{x^0} = \partial/ \partial{c(t)t}$ would be problematic. Instead we use $x^0$ replacing it by $c(t) t$ after performing all the differentiations.

\section{Varying constants quantum cosmology}
\label{Action}
\setcounter{equation}{0}

Classical cosmology of varying $c$ and $G$ theories have been studied widely and now we concentrate on quantizing it. The simplest quantization approach known as canonical quantization of gravity was proposed by DeWitt \cite{deWitt} and Wheeler \cite{Wheeler} and we follow the procedure. In order to get quantum Wheeler-deWitt equation we start with the Einstein-Hilbert action as in standard Refs. \cite{LL,Will} in the form
\begin{eqnarray}
\label{EHaction}
S_{g}=\int\limits_{M}d^{4}x \sqrt{-g}  \ ^{(4)}R \frac{c^3}{16\pi G} \label{action_grav}
\end{eqnarray}
where in general $c$ and $G$ may vary as $c=c(x^{\mu})$ and $G=G(x^{\mu})$, where $x^{\mu} = (x^0, x^1, x^2, x^3)$ and $x^0 = ct$. The action has dimension $J\cdot s$, the Ricci scalar $[R]= m^{-2}$, $[G]=Nm^2 kg^{-2}$, and the dimension of $d^{4}x \sqrt{-g} $ should be the 4- volume i.e. $m^4$. In Ref. \cite{harko2000} (and in many other papers - among them in \cite{VSL,VG}) the action (\ref{EHaction}) contains $c^4$ term instead of $c^3$. This is due to the fact that the authors there considered the coordinate $x_0 = t$ and not $x^0 = ct$ which can be concluded from Eq. (9) of Ref. \cite{harko2000} in which explicit dependence on $c$ was given.

%In Minkowski space:
%\be
%ds^2 = -(dx^0)^2+(dx^1)^2+(dx^2)^2+(dx^3)^2,
%&=& -c^2dt^2+(dx^1)^2+(dx^2)^2+(dx^3)^2,
%\ee
%$where $x^0= ct$.
In our approach we take the Friedmann metric
\be
\label{Fried}
ds_F^2= -(dx^0)^2+a^2(x^0)[d\chi^2+S^2(\chi)d\Omega^2]~,
\ee
where
\begin{eqnarray}
S(\chi)= \left\{ \begin{array}{ll}
\sin\chi, & \textrm{k=+1,}\\
\chi, & \textrm{k= 0,}\\
{\textrm{sh}}\chi,& \textrm{k= -1 }
\end{array} \right.
\end{eqnarray}
so that the dimension of $[d^4x_F=dx^0 dx^1 dx^2 dx^3]=m$, since the coordinates $x^1=\chi$, $x^2=\theta$, $x^3=\phi$ are dimensionless and $[\sqrt{-g_F}]=m^3$. This all gives $[d^4x\sqrt{-g_F}]=m^4$, so that the action (\ref{action_grav}), as it stands, has the right dimension $J\cdot s$. \\
According to Refs. \cite{LL,Will} the matter action reads as
\begin{equation}
S_m=\int\limits_{M}L_m\sqrt{-g}d^4x= -\int\limits_{M}\rho c\sqrt{-g}d^4x~, \label{action_matter}
\end{equation}
where $\varrho$ is the matter density, and the variation of (\ref{action_grav}) and (\ref{action_matter}) is
\begin{eqnarray}
 \frac{\delta S_g}{\delta g_{\alpha\beta}}&=&- \frac{c^3}{16\pi G}E^{\alpha\beta},\\
 \frac{\delta S_m}{\delta g_{\alpha\beta}}&=& \frac{1}{2c}T^{\alpha\beta},
\end{eqnarray}
which gives the Einstein equations in the form
\begin{equation}
E^{\alpha \beta}= R^{\alpha \beta}- \frac{1}{2}g^{\alpha \beta}R= \frac{8\pi G}{c^4} T^{\alpha \beta}. \label{Einstein_eq}
\end{equation}
Although some authors \cite{bojowaldbook} use different methods, we add Gibbons-Hawking boundary term
\begin{equation}
S_{GH}=\int\limits_{\partial M}d^3 x\sqrt{h}K \frac{c^3}{8 \pi G}, \label{action_GH}
\end{equation}
where $h$ is the determinant of a 3-metric, $K$ is extrinsic curvature $(K_{\mu\nu}=\frac{1}{2}h_{\mu\nu ,0} ,\  K_{11}=K_{22}=K_{33}=a,_{0}a, K= K_{\mu\nu}g^{\mu\nu}=3a,_{0}/{a}, \ [a,_{0}]= 1, [K]= m^{-1}$ and $[d^3x K]= m^3)$ and the dimension of (\ref{action_GH}) is also $J \cdot s$. As for the Einstein equations  (\ref{Einstein_eq})  we have the dimensions
\begin{equation}
\left[\frac{c^4}{G}\right]= N, \ \ [T_\alpha\   ^\beta] =\frac{J}{m^3} \nonumber
\end{equation}
and for the perfect fluid
\begin{equation}
T_\alpha\   ^\beta= \left( \varrho c^2+p \right)U_\mu U^{\nu} +p \delta _\mu^\nu
\end{equation}
we have $[\delta_\alpha ^{\beta} R]= m^{-2}$,  $[(8\pi G/c^4) T_\alpha\   ^\beta ]= m^{-2}$ as it should be.
For the matter in terms of the cosmological constant we have
\begin{equation}
S_\Lambda= -\int\limits_{M} \frac{2 \Lambda c^3}{16 \pi G} \sqrt{-g} d^4 x . \label{action_L}
\end{equation}
For the Friedmann metric (\ref{Fried}) we obtain
\begin{eqnarray}
\sqrt{-g}&=& a^3(x^0)S^2(\chi)\sin\theta,\nonumber\\
\sqrt{h}&=& a^3(x^0)S^2(\chi)\sin\theta,\nonumber\\
^{(4)}R&=& \frac{6}{a^2}\left( k+a^2,_0+aa,_{00} \right) , \label{FM}
\end{eqnarray}
and so
\begin{eqnarray}
\int\limits_{M}d^4 x\sqrt{-g}&=&\int a^3(x^0)dx^0 \int S^2(\chi)\sin\theta\ d\chi d\theta d\phi \nonumber\\
&=&V_3\int a^3(x^0)dx^0,
\end{eqnarray}
where the we have defined the volume factor
\begin{equation}
V_3= \int\limits_{M} S^2(\chi)\sin\theta d\chi d\theta d\phi. \label{objetosc}
\end{equation}
Obviously, for $ k=+ 1$ we have
\begin{displaymath}
V_3= \int\limits_{0}^{\pi} d\chi  \int\limits_{0}^{\pi} d\theta  \int\limits_{0}^{2\pi}d\phi \sin\chi \sin\theta= 2\pi^2,
\end{displaymath}
while for $k=0,-1$ we need a cut-off of the volume to have it finite.
The gravitational action (\ref{action_grav}) with the help of (\ref{FM})- (\ref{objetosc}) is given by
\begin{equation}
S_g=\frac{V_3}{16\pi}\int dx^0 a^3(x^0)\frac{c^3}{G}\left[\frac{6}{a^2(x^0)}\left( k+a^2,_0+aa,_{00} \right)\right] \label{a_g}
\end{equation}
and the boundary term is obtained as
\begin{eqnarray}
S_{GH}&=&-\frac{3}{8\pi}\int\limits_{\partial M} dx^0 a^3(x^0)\frac{c^3}{G} \frac{a,_0(x^0)}{a(x^0)} S(\chi) \sin\theta  \nonumber\\
&=&-\frac{3}{8\pi}\int\limits_{M}\frac{\partial}{\partial x^0} \left( a,_{0}a^2 \right) \frac{c^3}{G}S(\chi) \sin\theta  d^4 x \label{a_GH}
\end{eqnarray}
In fact, we vary the action in the special frame where $c$ and $G$ are constant while in the other frames they vary yielding the extra terms with the derivatives (cf. the discussion of Ref. \cite{EllisUzan03} about the proper usage of the variational principle for varying speed of light cosmology).

%{\bf Should be $\partial/\partial x^0 (a,0 a^2 c^3)/G$ and killed with extra terms of the type $\delta c$ and $\delta G$ in the EH action. Alternatively we could neglect %the total derivative wrt time - not adding the boundary term $S_{GH}$.} If we varied the action (\ref{EHaction}) with respect to $c$ and $G$, then the boundary term could %have been taken to have them varying too and these both terms would cancel .

For the cosmological term we have:
\begin{eqnarray}
S_{\Lambda}= -\frac{3 V_3}{8\pi}\int \frac{\Lambda c^3}{3G a^3(x^0)}dx^0 \label{a_L} ,
\end{eqnarray}
while for the matter term:
\begin{eqnarray}
S_m&=&-\int  \varrho(x^0) c \sqrt{-g}dx^0\nonumber\\
&=&-V_3\int  \varrho(x^0)c a^3(x^0)dx^0 \label{a_m}.
\end{eqnarray}
Collecting all the terms (\ref{a_g})-(\ref{a_m}) we have the total action and apply the variability of $c$ and $G$ here
\begin{eqnarray}
&& S= S_g+S_{GH}+S_{\Lambda}+ S_m = \frac{3V_3}{8\pi}\int dx^0 \frac{c^3(x^0)}{G(x^0)} \times \nonumber \\
&& \left[ka- a,_0^2a-\frac{\Lambda}{3}a^3-\frac{8\pi G(x^0)}{3c^2} \varrho a^3 \right]. \label{action}
\end{eqnarray}
%Using the ''t'' coordinate instead of $x^0$ one has $da/dx^0=da/cdt=\dot{a}/c$ \footnote{However, there is a problem since formally $c=c(t)$ and so this derivation is %valid provided we assume $c$ constant.} and so we have
%\begin{equation}
%S= \frac{3V_3}{8\pi}\int dt\frac{c^2(t)}{G(t)}\left[kac^2- \dot{a}^2a-\frac{\Lambda}{3}a^3c^2-\frac{8\pi G(t)}{3} \varrho a^3 \right]
%\end{equation}
According to (\ref{action}) the Lagrangian reads as
\begin{equation}
L=\frac{3V_3 c^3(x^0)}{8\pi G(x^0)}\left(ka- a,_0^2a-\frac{\Lambda}{3}a^3-\frac{8\pi G(x^0)}{3c^2} \varrho a^3 \right)\,
\end{equation}
so that the conjugate momentum is \cite{deWitt})
\begin{equation}
p_a=\frac{\partial{L}}{\partial{a,_0}}= -\frac{3V_3c^3}{4\pi G} a a,_0 \,
\end{equation}
and the Hamiltonian reads as
\begin{eqnarray}
\label{hamilt}
H&=& p_a a,_0- L = -\frac{2\pi G(x^0)}{3V_3 c^3(x^0) a} p_a ^2 ,  \\
&-& \frac{3V_3 c^3(x^0)}{8\pi G(x^0)} k a + \frac{V_3c^3(x^0)}{8\pi G(x^0)} \Lambda a^3 +V_3 \varrho c(x^0) a^3 . \nonumber
\end{eqnarray}
In order to quantize (\ref{hamilt}) we replace $p_a$ for the operator
\begin{equation}
p_a\rightarrow -i\hslash \frac{\partial}{\partial{a}} ,
\end{equation}
and choosing the simplest factor ordering \cite{vilenkin} we have Wheeler-DeWitt equation in the form
\begin{equation}
\left[\hslash^2\frac{\partial ^2}{\partial{a^2}} - U(a)\right] \Psi(a)= 0 ,
\end{equation}
where the minisuperspace (only one-dimensional with scale factor $a$ as variable) potential reads as
\begin{eqnarray}
\label{pot1}
&& U(a)= -\left(\frac{3V_3 c^2(a)a}{4\pi G(a)}\right)^2 \times \nonumber \\
&& \left[ kc^2(a)  - \frac{\Lambda}{3} a^2c^2(a) -\frac{8\pi G(a)}{3} \varrho(a) a^2 \right].
 \label{potential}
\end{eqnarray}
In (\ref{pot1}) we replaced the dependence of $c$ and $G$ on time $x^0 = ct$ by the dependence on the scale factor, since in the minisuperspace there is no classical time and the role of the ``time'' parameter can be played by the scale factor $a$ \cite{deWitt,kieferbook}. After imposing the barotropic equation of state $p=w\varrho c^2$ with $w=$ const. and using the ans\"atze (\ref{c(t)}) and (\ref{G(t)}) we have
\bea
\label{pot2}
&& U(a)= -\left(\frac{3V_3 c_0^2 a^{2n+1-q}}{4\pi G_0}\right)^2 \times \nonumber\\
&& \left( kc_0^2a^{2n}  - \frac{\Lambda}{3} c_0^2a^{2n+2} -\frac{8\pi G_0}{3} a^{2+q}  \varrho(a)\right),
\eea
and after inserting (\ref{varrho}) to the potential (\ref{pot2}), one has
\begin{eqnarray}
\label{potential2}
&& U(a)= -K_0^2 a^{2(3n+1-q)}  \\
&& \left( \frac{3w+1}{2n+3w+1}k - \frac{\Lambda(w+1)}{2n+3(w+1)}a^2 -\frac{8\pi G_0}{3c_0^2}\frac{C}{a^{3w+1+2n}} \right), \nonumber
\end{eqnarray}
where we have defined
\begin{equation}
K_0=\frac{3V_3 c_0^3 }{4\pi G_0} .
\end{equation}
It is easy to notice that (\ref{potential2}) gives Eq. (7) of Ref. \cite{MithaniSHU} for $n=q=0$ and $V_3 = 2\pi^2$.
Assuming that $C=0$ in which case the dependence of the minisuperspace potential on the matter content is only due to dissipative term on the right-hand side of the conservation equation (\ref{conserLam}) related to variability of $c$ and vanishes if $n=0$. Under this assumption the potential (\ref{potential2}) reads as
\begin{equation}
\label{potential3}
U(a)= K_0^2 a^{2(3n+1-q)} \left( \frac{(3w+1)k}{2n+3w+1} - \frac{\Lambda(w+1)a^2}{2n+3(w+1)} \right),
\end{equation}
and it has zeros at $a=0$ as well as at
\begin{eqnarray}
\label{zero}
a_t=\sqrt{\frac{k(3w+1)[2n+3(w+1)]}{ \Lambda(w+1)(2n+3w +1)}}
\end{eqnarray}
provided $a_t$ is a real number. For radiation ($w=1/3$) we recover the Eq. (67) of Ref. \cite{szydloQC}. In Ref. \cite{harko2000} the shape of the potential differs from ours. The point is that in this reference the conservation equation (\ref{conserLam}) was not integrated explicitly. Instead, a specific ansatz for the mass density $\varrho = \varrho_0 a ^n$ similar to our ans\"atze for $c$ and $G$ in (\ref{c(t)})-(\ref{G(t)}) was assumed and checked by the conservation equation for consistency. This differs from our approach.

 \begin{figure}
 %[ht]
 %\unitlength1cm
 \begin{center}
 \scalebox{0.8}{\includegraphics[angle=0]{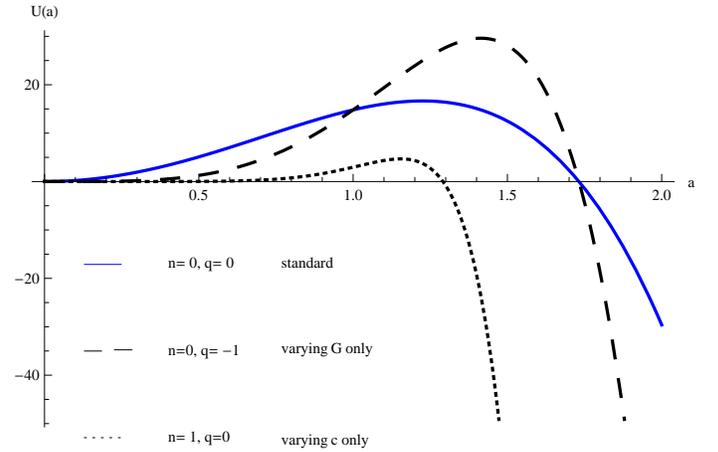}}
\caption{The potential (\ref{potential3}) in minisuperspace for dust $w=0$ and $\Lambda =1$ models which allow the tunneling of the universe from $a=0$ to $a=a_t$. Three cases are shown: the standard one with the cosmological term only (cf. Ref. \cite{vilenkin86,atkatz}) and with no variation of $c$ and $G$ (solid blue line); the varying $c$ model (dotted line), and the varying $G$ model (dashed line). The zero (\ref{zero}) of the potential (\ref{potential3}) does not depend on the variability of $G$, but varying $G$ makes the potential less steep near to $a=0$ and rises the height of the barrier. In varying $c$ models the height of the barrier and the size of the created universe are smaller than in the standard case.}
\label{fig1}
 \end{center}
 \end{figure}

\begin{figure}
 %[ht]
 %\unitlength1cm
 \begin{center}
 \scalebox{0.8}{\includegraphics[angle=0]{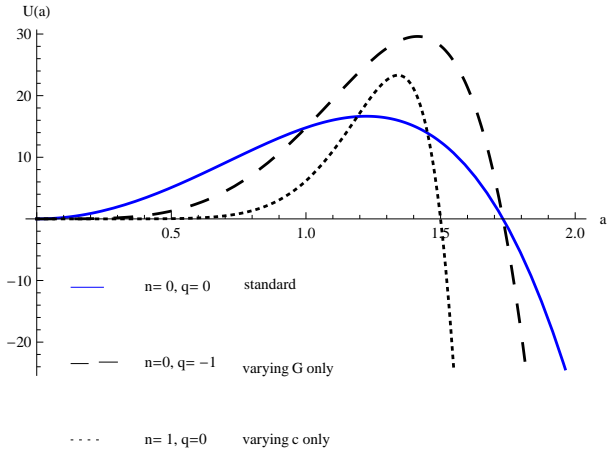}}
\caption{The potentials (\ref{potential3}) in minisuperspace for radiation $w=1/3$ and $\Lambda =1$ models which allow the tunneling of the universe from $a=0$ to $a=a_t$. Three cases are shown: the standard one with the cosmological term only (cf. Ref. \cite{vilenkin86,atkatz}) and with no variation of $c$ and $G$ (solid blue line); the varying $c$ model (dotted line), and the varying $G$ model (dashed line). As in Fig.\ref{fig1} the zero of the potential (\ref{zero}) does not depend on the variability of $G$, but varying $G$ makes the potential smoother near to $a=0$ and rises the height of the barrier. In varying $c$ models the height of the barrier is higher and the size of the created universe is smaller than in the standard case.}
\label{fig2}
 \end{center}
 \end{figure}

In Figs. \ref{fig1} and \ref{fig2} the plots for the minisuperspace potential (\ref{potential3}) for $\Lambda = 1$ are given. All of them are of the tunneling type. Both varying $c$ and varying $G$ potentials are not as steep near to the $a=0$ as in the standard potential with the cosmological term only, as given in Refs. \cite{vilenkin86,atkatz}. The height of the barrier is larger than in the standard case for dust $(w=0)$ varying $G$ models (cf. Fig. \ref{fig1}), while it is higher for both varying $c$ and varying $G$ for radiation $(w=1/3)$ models. We have checked that the same is true for stiff-fluid models.

 \begin{figure}
 %[ht]
 %\unitlength1cm
 \begin{center}
 \scalebox{0.8}{\includegraphics[angle=0]{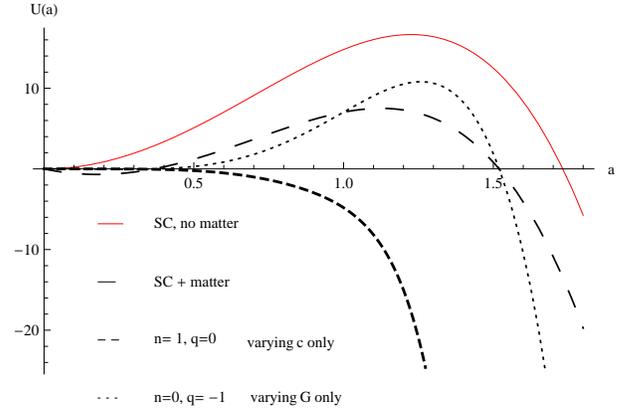}}
\caption{The potential (\ref{potential2}) in minisuperspace for $w=0$, and $\Lambda=1$ models. Four cases are shown: the standard one with the cosmological term only (cf. Ref. \cite{vilenkin86,atkatz}) for $C=0$ and no variation of $c$ and $G$ (solid line); the standard $\Lambda$-term model with $C = 0.35$ and dust $w=0$ (dot-dashed lines); the varying $c$ model (dashed line); and the varying $G$ model (dotted line). It is interesting that for varying $c$, the potential is always negative which means that there are no classically forbidden regions for $U>0$, while for the standard $C \neq 0$ case it has two classically allowed regions: one which is oscillating (the universe starts from a singularity expands and reaches maximum, then recollapses) and another which is asymptotic (starts from a finite size and then expands forever). This implies a possibility for the universe to tunnel at the maximum expansion point to an asymptotic regime (to an asymptotic model to the Einstein static universe).}
\label{fig3}
 \end{center}
 \end{figure}

\begin{figure}
 %[ht]
 %\unitlength1cm
 \begin{center}
 \scalebox{0.8}{\includegraphics[angle=0]{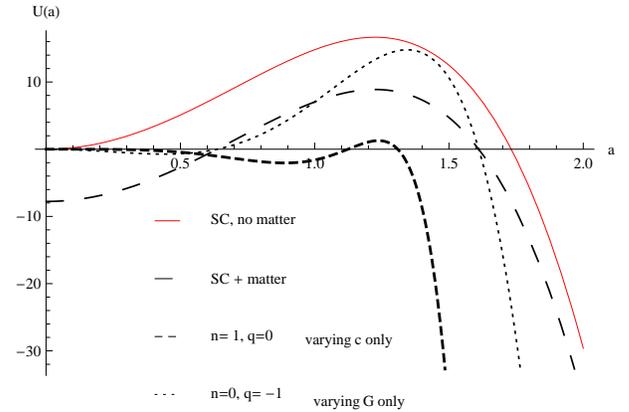}}
\caption{The potential (\ref{potential2}) in minisuperspace for $C = 0.35$, $w=1/3$, and $\Lambda=1$ models. Four cases are shown: the standard one with the cosmological term only (cf. Ref. \cite{vilenkin86,atkatz}) for $C=0$ and no variation of $c$ and $G$ (solid line); the standard $\Lambda$-term model (dot-dashed line); the varying $c$ model (dashed line); and the varying $G$ model (dotted line). For a standard pure radiation and $\Lambda$ model there is also a classically allowed region near $a=0$.}
\label{fig4}
 \end{center}
 \end{figure}

In Figs. \ref{fig3} and \ref{fig4} the plots for the minisuperspace potential (\ref{potential2}) for $C \neq 0$ and $\Lambda = 1$ models are shown. For dust $w=0$ models (cf. Fig. \ref{fig3}) most of them are of the tunneling type except for varying $c$ models for which the potential is of the scattering type (there is no classically forbidden region for $U>0$). For radiation $w=1/3$ models (cf. Fig. \ref{fig4}) there is an extra classically allowed region between $a=0$ and some $a=a_{max}$ for the potential which allows oscillations (the universe starts from a singularity expands and reaches maximum, then recollapses) as well as the asymptotic expansion from some $a_b$ to infinity (starts from a finite size and then expands forever). This implies a possibility for the universe to tunnel at the maximum expansion point to an asymptotic regime (compare Refs. \cite{D+L,ANN96,MithaniSHU,SHU,instabSHU,graham14,stabSHU} where the tunneling is possible to an oscillating regime). For a standard pure radiation and $\Lambda$ model there is also a classically allowed region near $a=0$.

\section{Quantum tunneling in varying constants cosmology}
\setcounter{equation}{0}
\label{tunnel}

Now, we use the WKB method \cite{vilenkin86,atkatz} to calculate the probability of tunneling of the universe ``from nothing'' ($a=0$) to a Friedmann geometry with $a=a_t$ which reads as
\begin{eqnarray}
\label{probab1}
P&\simeq&\exp\left[ -\frac{2}{\hslash}\int\limits_0^{a_t} \sqrt{2(E-U(a))}da\right]\nonumber\\
&=&\exp\left[-\frac{2K_0}{\hslash} \int\limits_0^{a_t}a^{3n+1-q}\left( \frac{\Lambda(w+1)}{2n+3(w+1)}a^2 \right. \right. \nonumber \\
&-& \left. \left. \frac{3w+1}{2n+3w+1}k\right)^{1/2} da \right].
\end{eqnarray}
In (\ref{probab1}) we have assumed the simpler potential (\ref{potential3}) for $C=0$ since for $C \neq 0$ the integral is not analytic. Such an integral can of course be integrated numerically.

Making the substitution $a^2= \tilde x$, we have
\begin{eqnarray}
P&\simeq&\exp\left[-\frac{K_0}{\hslash} \int\limits_0^{\tilde x}\tilde x^{\frac{1}{2}(3n-q)} \left( \frac{\Lambda(w+1)}{2n+3(w+1)}\tilde x \right. \right. \nonumber \\
&-& \left. \left. \frac{3w+1}{2n+3w+1}k\right)^{1/2}d\tilde x\right].
\end{eqnarray}
Putting $\tilde x=\tilde x_0x$, where
\begin{equation}
\tilde{x}_0=\frac{k(3w+1)(2n+3w+3)}{\Lambda(2n+3w+1)(w+1)}
\end{equation}
one has
\begin{eqnarray}
&& P\simeq\exp\left[ -\frac{K_0}{\hslash}\sqrt{\left| \frac{3w+1}{2n+3w+1}k\right|}\ \tilde{x}_0^{\frac{1}{2}(3n-q+2)} \right. , \nonumber \\
&& \left. \times \int\limits_0^1x^{\frac{1}{2}(3n-q)}(1-x)^{1/2}dx\right],
\end{eqnarray}
with an additional condition that $3n+2 > q$.
%Comparing this with the definition of Beta function
%\begin{equation}
%B(p,r)= \int\limits_0^1 x^{p-1}(1-x)^{r-1} dx= \frac{\Gamma(p)\Gamma(r)}{\Gamma(p+r)}
%\end{equation}
%one can get
%\begin{eqnarray}
%p&=&\frac{1}{2}(3n+2-q)\nonumber\\
%r&=&\frac{3}{2},\nonumber
%\end{eqnarray}
which after using the definition of Beta and Gamma functions \cite{abramovitz} gives the tunneling probability
\begin{eqnarray}
P&\simeq&\exp\left[ -\frac{K_0}{\hslash}\sqrt{\left| \frac{3w+1}{2n+3w+1}k\right|} \right. \nonumber \\
&& \left. \tilde{x}_0^{\frac{1}{2}(3n-q+2)}B\left( \frac{1}{2}(3n+2-q), \frac{3}{2}\right)\right]\nonumber\\
&\simeq&\exp\left[ -\frac{K_0}{\hslash}\sqrt{\left| \frac{3w+1}{2n+3w+1}k\right|} \right. \nonumber \\
&& \left. \tilde{x}_0^{\frac{1}{2}(3n-q+2)} \frac{\sqrt{\pi}}{2}\frac{\Gamma\left( \frac{1}{2}(3n+2-q)\right)}{\Gamma\left( \frac{1}{2}(3n+5-q)\right)}\right]. \label{probability}
\end{eqnarray}

  \begin{figure}[ht]
 %\unitlength1cm
 \begin{center}
 \scalebox{0.8}{\includegraphics[angle=0]{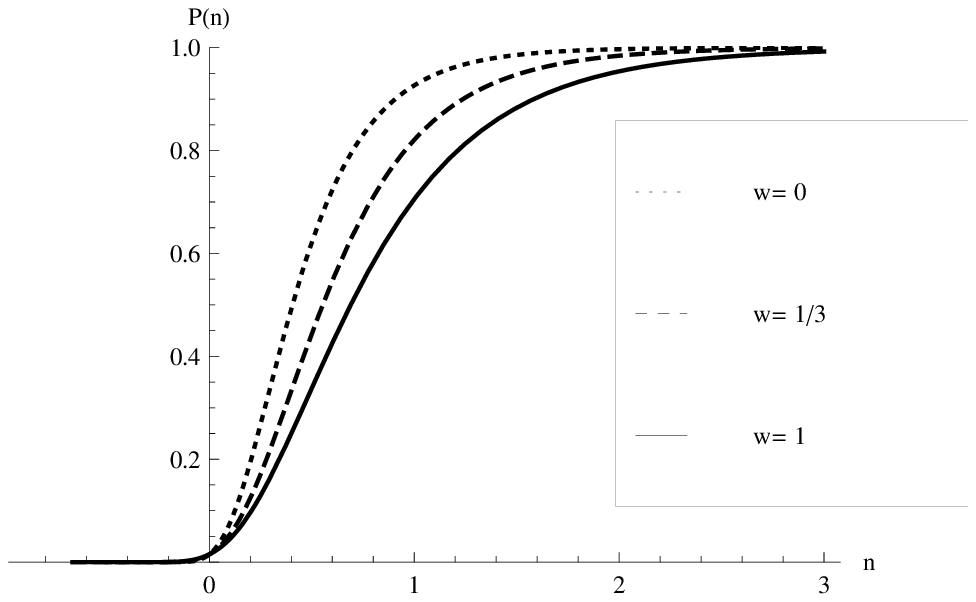}}
 \scalebox{0.6}{\includegraphics[angle=0]{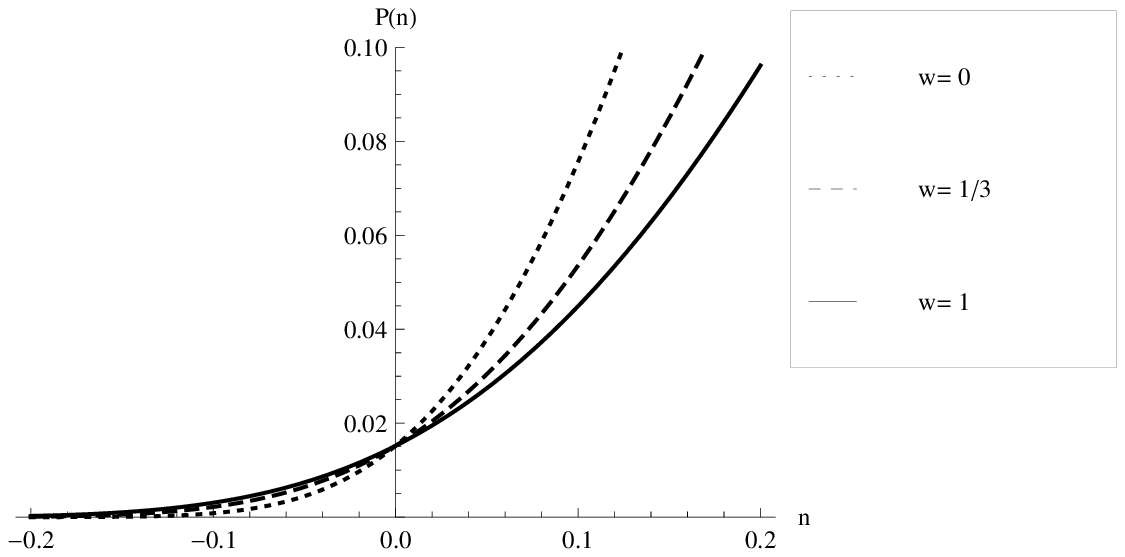}}
\caption{The probability of tunneling (\ref{probability}) for varying $c$ minisuperspace models with $k=+1$, $q=0$, $\Lambda=5$, and $\hslash = G_0 = c_0 = 1$. Three cases are shown: dust ($w=0$), radiation ($w=1/3$), and stiff-fluid ($w=1$). The probability of tunneling for the negative values for the parameter $n$ (diminishing speed of light $c$) is very small (below) in comparison to its positive values, where it rises sharply and then asymptotes to one (above). The probability is smallest for the stiff fluid though it asymptotes towards one, too.}
\label{Pn}
 \end{center}
 \end{figure}

\begin{figure}[ht]
 %\unitlength1cm
 \begin{center}
 %\scalebox{0.8}{\includegraphics[angle=0]{3_P(q).eps}}
 \scalebox{0.9}{\includegraphics[angle=0]{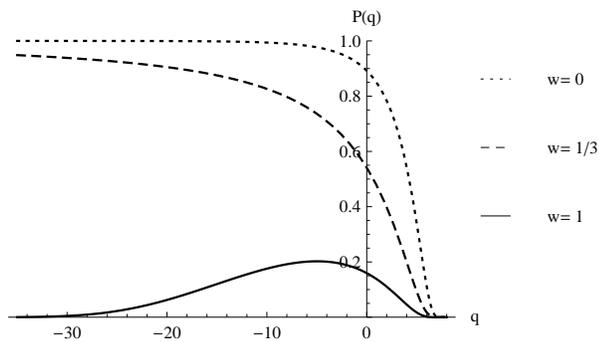}}
\caption{The probability of tunneling (\ref{probability}) for varying $G$ minisuperspace models with $k=+1$, $n=2$, $\Lambda=2$, and $\hslash = G_0 = c_0 = 1$. Three models are shown: dust ($w=0$), radiation ($w=1/3$), and stiff-fluid ($w=1$). The probability of tunneling for the negative values of the parameter $q$ is larger than for its positive values where it drops to zero for a certain value of $q$.}
\label{Pq}
 \end{center}
 \end{figure}

We present the plots for the probability of tunneling (\ref{probability}) of the universe ``from nothing'' ($a=0$) to a Friedmann universe with some value of the scale factor $a_t$ for varying $c$ and $G$ models in Figs.\ref{Pn} and \ref{Pq}. The probability of tunneling is very small for negative values of the parameter $n$ which corresponds to decreasing speed of light $c$, reaches the value of about $0.015$ for $n=0$ and further increases up to one for positive values of the parameter $n$ which corresponds to increasing speed of light $c$. The plot in Fig.\ref{Pn} shows that the claim of Ref. \cite{szydloQC} that the probability of tunneling was largest for a constant value of the speed of light $(n=0)$ was true only if one admitted the decrease of the speed of light $(n<0)$ which was based on the first observations of the variability of the fine structure constant $\alpha \propto c^{-1}$ \cite{webb99,murphy2007}. However, in view of new observational results both the decrease and the increase of $c$ are reported \cite{webb2011,king2012}, so that one should not restrict oneself to such a case only. As we see from Fig.\ref{Pn}, for speed of light increasing, the probability of tunneling is monotonically increasing either, and no maximum appears. On the other hand, from Fig.\ref{Pq} we conclude that the probability of tunneling for the negative values of the parameter $q$ (diminishing gravitational constant $G$ or weakening of gravity) is larger than for its positive values, where it drops to zero for a certain value of $q$. For stiff-fluid, the probability is suppressed both for positive values of $q$ as well as for its large negative values.

\begin{figure}[ht]
 %\unitlength1cm
 \begin{center}
 \scalebox{0.9}{\includegraphics[angle=0]{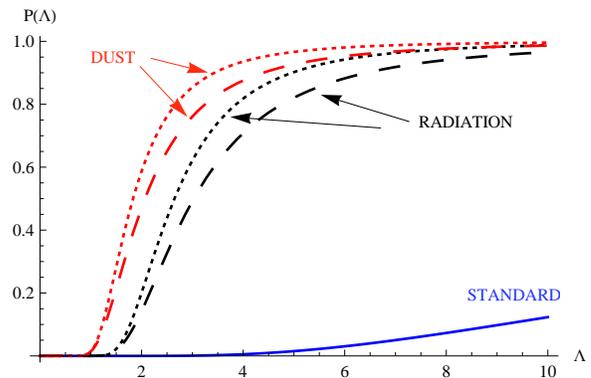}}
\caption{The probability of tunneling (\ref{probability}) versus the cosmological constant $\Lambda$ with $k=+1$ and $\hslash = G_0 = c_0 = 1$. The lowest solid blue line is for the standard cosmological term minisuperspace potential only (cf. Ref. \cite{vilenkin86,atkatz}). The dark lines are for radiation (lower ones) and red lines are for dust (higher ones). The dashed lines are for varying $c$ only ($n=1$, $q=0$), while the dotted lines are for both $c$ and $G$ arying models ($n=1$, $q=-1$). The probability is highest for dust and also it is larger for both $c$ and $G$ varying.}
\label{PL}
 \end{center}
 \end{figure}

In Fig.\ref{PL} the probability of tunneling is plotted against the cosmological constant. From the plot we conclude that the probability of tunneling is growing proportionally to the value of the (positive) cosmological constant and that its growth is higher if one of the types of matter (dust, radiation) are present. Also, it is clear that both the variability of $c$ and $G$ rise the probability of tunneling under the specific choice of the parameters of our model ($n=1$, $q=-1$).

%{\bf If one has not neglected the integration constant in (\ref{varrho}) the potential takes the form given by the equation (\ref{potential2}). This form can be compared %with Mithani \& Vilenkin (1204.4658) formulas (7) and (15) (here $\varrho(a)= \Lambda + \varrho_0/a$ for $w_\Lambda= -1$ and $w_0= -2/3$)..... the one considered in Refs. %\cite{ANN96,D+L} and later in \cite{MithaniSHU} where the oscillating classical models with domain walls and negative $\Lambda$ has been considered and allowed to tunnle %between oscillating and expanding/singular regimes. In most of our cases with varying $c$ and $G$ there exists non-zero probability for the universe to collapse onto a %singularity \cite{SHU,instabSHU,graham14,stabSHU}.\\
%The WKB integration will be more difficult in view of the first term in (\ref{potential2}) except for some special cases like phantom with $w=- 5/3$ which gives the term %$\sim a^4$.}

%{\bf Two interesting physically cases can be studied: first when we have radiation matter ($w = 1/3$) and dust matter ($w = 0$), positive curvature $k=+1$ and positive %cosmological constant $\Lambda > 0$. Second when we have the network of domain walls ($w = -2/3$), negative curvature $k=-1$ and positive cosmological constant.}

\section{Conclusions}
\setcounter{equation}{0}
\label{conclusion}

We have discussed quantum cosmology of varying speed of light $c$ and varying gravitational constant $G$ theories with matter terms and including the cosmological constant. We followed the standard canonical quantization approach and after detailed discussion of the action and boundary terms for such theories we obtained the corresponding Wheeler-DeWitt equation. We have found that most of the varying $c$ and $G$ minisuperspace potentials are of the tunneling type. This allowed us to use WKB approximation of quantum mechanics to calculate the probability of tunneling of the universe ``from nothing'' i.e. from the initial singular state at $a=0$ to a Friedmann universe with isotropic geometry characterized by some fixed value of the scale factor $a_t$. We have obtained that the probability of tunneling depends on both $c$ and $G$ being not constant. In fact, it is large for growing $c$ models and is strongly suppressed for diminishing $c$ models. On the other hand, the probability of tunneling is large for gravitational constant $G$ decreasing (i.e. for weaker gravity), while it is small for $G$ increasing. In general, the variability of $c$ and $G$ influences the probability of tunneling as compared with the standard matter content universe models.

An interesting feature of varying speed of light cosmology is that the probability of tunneling depends directly on the matter content through the dissipative terms on the right-hand side of the conservation equation (\ref{conserLam}) despite the fact that the form of the Einstein equations is not changed (which is a consequence of the fact that one makes specific variation of the gravitational action which is valid only in a special frame). In our model this means that the integration constant $C$ is zero in such a case. Obviously, if the speed of light is constant $\dot{c}=0$, then $n=0$ in (\ref{varrho}) and we get the standard quantum cosmology case. However, for $C \neq 0$ the matter content enters the minisuperspace potential in a standard way and we have the probability of tunneling influenced by the matter content both ways: by the dissipative terms and by the standard field equations.

\section{Acknowledgements}

MPD would like to thank Remo Garratini, Marek Szyd{\l}owski, and Alex Vilenkin for discussions. This project was financed by the National Science Center Grant DEC-2012/06/A/ST2/00395.

%\appendix

%\section{Choice of units/ Derivation of varying constants minisuperspace equations starting from the Eintein-Hilbert action}
%\label{appendixA}
%\setcounter{equation}{0}

\end{document}